\documentclass[12pt]{article}
\usepackage{color}
\usepackage{amsmath}
\usepackage{graphicx}
\usepackage{setspace}

\usepackage[top=1.54cm,bottom=2.54cm,left=2.54cm,right=2.54cm]{geometry}
\begin{document}

\begin{titlepage}

\title{Color-Kinematic Duality in ABJM Theory Without Amplitude Relations}
\author{Allic Sivaramakrishnan
\\
\\ \emph{Department of Physics and Astronomy,}
\\ \emph{University of California, Los Angeles}
\\ \emph{Los Angeles, CA 90095-1547, USA}
\\
\\ \emph{allic@physics.ucla.edu}}

\date{}

\maketitle	

\begin{abstract}
We explicitly show that the Bern-Carrasco-Johansson color-kinematic
duality holds at tree level through at least eight points in
Aharony-Bergman-Jafferis-Maldacena theory with gauge group SU(N)
x SU(N). At six points we give the explicit form of numerators
in terms of amplitudes, displaying the generalized gauge freedom
that leads to amplitude relations.  However, at eight points no amplitude
relations follow from the duality, so the diagram numerators are fixed
unique functions of partial amplitudes. We provide the explicit
amplitude-numerator decomposition and the numerator relations for
eight-point amplitudes.
\end{abstract}
	
\end{titlepage}

\section{Introduction}
Studies of scattering amplitudes have uncovered important insights
into gauge and gravity theories. In particular, Bern, Carrasco, and
Johansson (BCJ) found a surprising duality between the color
factors and kinematic numerator factors that comprise diagrams in Yang-Mills theory ~\cite{BCJ}. 
The color-kinematic duality implies nontrivial amplitude relations. 
These amplitude relations have been studied in both field theory and string
theory~\cite{Tye10, Stieberger10, Bjerrum-BohrDSV10}. The color-kinematic duality appears
to extend to loop level, as confirmed in a variety of examples with
varying levels of supersymmetry~\cite{BCJLoop,LoopExamples}.
While there has been some progress in understanding the origin of the
duality from a Lagrangian vantage point~\cite{BernDHK10, Monteiro11,
  Tolotti13}, further work is needed. 

The color-kinematic duality reveals new structures in gravity through
a surprisingly simple gauge-gravity correspondence. Kinematic
numerators satisfying the color-kinematics duality provide the link:
by replacing color factors with numerators that satisfy the duality,
gauge-theory amplitudes are converted into gravity amplitudes
\cite{BCJ,BCJLoop}, revealing a double-copy structure of gravity. A
connection between gravity and Yang-Mills theory has long been known
at tree level from the Kawai-Lewellen-Tye relations \cite{KLT}, but
the double-copy property of gravity reveals a more extensive
correspondence, one that appears to extend to loop level.  This
property has advanced the study of supergravity's
properties, especially in uncovering unexpected ultraviolet
cancellations at high loop orders~\cite{GravityUV}.  More generally,
the double-copy relation allows us to directly study the effects of
any newly uncovered properties of gauge-theory amplitudes on
corresponding gravity amplitudes.

In Yang-Mills theory, BCJ amplitude relations come from a residual
generalized gauge freedom present in kinematic numerators even after
imposing that the duality between color and kinematics is
manifest~\cite{BCJ,Bjerrum-BohrDSV10,Vaman10,Boels12,Litsey13}. There
have also been string-theory studies to investigate the residual gauge
invariance in the duality between color and kinematics in Yang-Mills
theory \cite{Tye10,Stieberger10}. While there has been
progress in understanding the underlying structure behind the duality
and the residual gauge invariance~\cite{Bjerrum-BohrDV10,Monteiro11},
further clarification is needed. To gain additional insight,
it is important to study a wide variety of cases where the 
duality holds.

In particular, the color-kinematic duality has been found in
three-dimensional Chern-Simons-matter theories: the $\mathcal{N}=8$
Bagger-Lambert-Gustavsson (BLG) theory and the $\mathcal{N}=6$
Aharony-Bergman-Jafferis-Maldacena (ABJM) theory
\cite{Bargheer12,HuangJ12}.  BLG theory turns out to be a special
case of ABJM theory.  These cases are quite interesting because a Lie
three-algebra, not a Lie two-algebra, defines the gauge structure of
these theories. These Chern-Simons-matter theories would appear to have
rather different properties than gauge theory. The color-kinematic
duality is governed by gauge-group relations, so the presence of the
duality in ABJM-type theories shows that the duality is more general
than previously appreciated.

We address the color-kinematic duality in ABJM theory. 
The first nontrivial example of the duality in this theory was first given in 
ref. \cite{Bargheer12}: the six-point all-scalar amplitude. While this
manuscript was in preparation, ref. \cite{HuangJL13} noted that the
duality holds up to ten points, but surprisingly found no BCJ amplitude
relations at eight points or higher. Here we provide specifics about
the eight-point amplitudes. The eight-point process is the simplest
example in which the color-kinematic duality holds without the
residual freedom that produces BCJ amplitude relations, warranting 
a detailed study.

Another curious feature that may be connected is that at lower
points, the double-copy property holds~\cite{Bargheer12,HuangJ12}, but
starting at eight points it does not~\cite{HuangJL13}:
applying the double-copy procedure to the eight-point kinematic numerators does not produce
the appropriate gravity numerators. The eight-point ABJM amplitude is
the only known instance in which the color-kinematic duality holds but
the double-copy property fails. Examining the eight-point amplitudes in 
some detail may therefore further illuminate the connection between the color-kinematic
duality and the double-copy property.

In this note, we start by presenting relatively explicit forms of the six point
numerators in ABJM theory that satisfy the duality between color and
kinematics.  This has not been given previously and is the first
nontrivial case where the duality holds. (At four points the duality
is trivial and for odd points the amplitudes vanish.)  We then confirm
that the duality between color and kinematics holds at eight points
for bosonic external states, but does not generate BCJ amplitude
relations. In an ancillary online file~\cite{Online}, we present a set
of independent eight-point amplitudes in terms of numerators, and the
numerator relations implied by the color-kinematic duality.

This note is organized as follows. In Sec. 2, we review the necessary
background. Sec. 2.1 describes the color-kinematic duality in
Yang-Mills theory. Then, in Sec. 2.2 we explain the construction of
partial amplitudes and how the color-kinematic duality leaves behind a
residual freedom that gives rise to BCJ amplitude relations. Sec. 2.3
presents relevant properties of the three-algebra formulation of
ABJM theory. Sec. 3 contains our results: demonstration of the
color-kinematic duality through eight points. In Sec. 3.1, we show the
trivial case: four points. We demonstrate the first non-trivial case
-- six points -- and present numerator solutions in terms of
amplitudes in Sec. 3.2. In Sec. 3.3 we detail the eight-point case. We
explain how to construct the eight-point partial amplitudes from
numerators, and confirm that the color-kinematic duality is satisfied
but does not imply BCJ amplitude relations. In Sec. 4 we discuss the
implications of the eight-point result and future directions.

\section{Review}

\subsection{The color-kinematic duality in Yang-Mills theory}
In Yang-Mills theory, tree amplitudes can in general be written as
\begin{equation}
\mathcal{A}(1,2,\ldots,n) = \sum_i  \frac{c_i n_i}{\prod_{\alpha_i} {s_{\alpha_i}}},
\label{eq:colornumdecomp}
\end{equation}
where the sum runs over diagrams with only cubic vertices. In general, any terms not 
of this form can be put into this form by multiplying and dividing by 
appropriate propagators. Here,
$c_i$ are products of structure constants and we suppress the coupling
constant and helicity labels. The $n_i$ are kinematic numerators:
functions of momenta and polarization vectors. Each color factor $c_i$
is in one-to-one correspondence with a diagram with a specific
propagator structure. The $s_{\alpha_i}$ in the denominator are the
Feynman propagators for the $i$-th diagram, where $s_{\alpha_i}$ are the
kinematic invariants of the scattering process.

\begin{figure}[tb]
\begin{center}
\includegraphics[width=\textwidth]{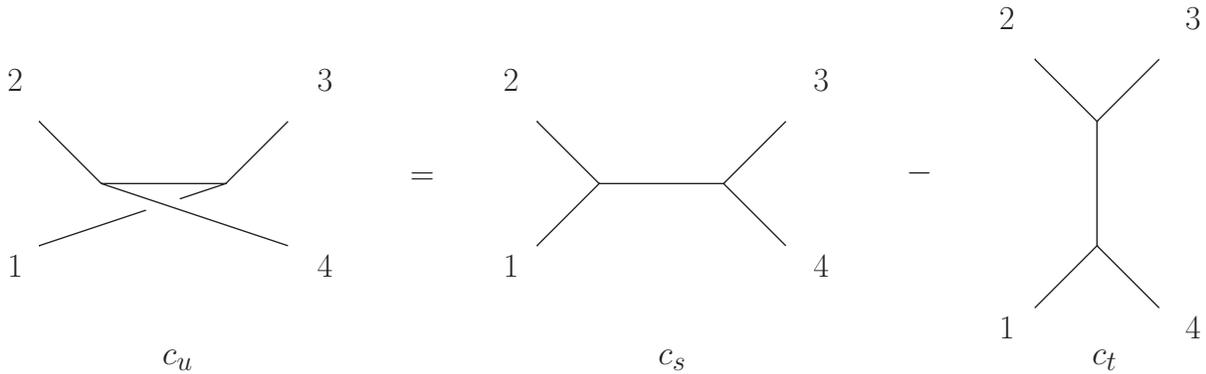}
\end{center}
\caption{\small 
Diagrammatic representation of the Jacobi relation.
}
\label{YMJacRelationFigure}
\end{figure}

The $c_i$ are not an independent set and are
related by Jacobi identities, as shown in fig.~\ref{YMJacRelationFigure}. BCJ proposed the
color-kinematics duality, wherein the numerators $n_i$ obey the same
relations and symmetries under relabelling \cite{BCJ}. For instance,
\begin{align}
c_1 = c_2 + c_3 &\Rightarrow n_1 = n_2 + n_3.
\\
c_1 \rightarrow -c_1 &\Rightarrow n_1 \rightarrow -n_1.
\end{align}
BCJ also noted that replacing $c_i$ with $n_i$ that satisfy the duality in
\eqref{eq:colornumdecomp} gives the scattering amplitude
$\mathcal{M}(1,2,\ldots,n)$ in gravity. This connection between gravity
and gauge theory is known as the double-copy property of gravity.
 At tree level, these properties have been proven in various
ways ~\cite{BernDHK10,Bjerrum-BohrDSV10,OtherDoubleCopyProofs}. 
The color-kinematic duality is a basic property of gauge and gravity 
theories that deserves further study, especially to understand the underlying symmetry.

\subsection{Amplitude relations from the color-kinematic duality in Yang-Mills theory}

The simplicity of the color-kinematic duality and double-copy property
suggests a novel principle in gauge theories and gravity.  
Here, we review how amplitude relations follow 
from the color-kinematic duality in Yang-Mills theory. We also note
that such amplitude relations arise from 
string theory \cite{Stieberger10,Bjerrum-BohrDV10}. The color-kinematic 
duality has also been directly studied in string theory~\cite{Tye10}. 

In Yang-Mills theory, we can write the color-decomposition in
\eqref{eq:colornumdecomp} in terms of color-ordered amplitudes and
traces over gauge-group generators. At tree level, this trace
decomposition is~\cite{ManganoReview}
\begin{equation}
{\cal A}(1,2,\ldots,n) =  \sum_{\sigma \in S_n/Z_n}{\text{Tr}(T^{\sigma_1}T^{\sigma_2}\ldots T^{\sigma_n}) A(\sigma_1, \sigma_2,\ldots \sigma_n)},
\end{equation}
where $S_n/Z_n$ is all leg orderings unrelated by cyclic
permutation. The color-ordered amplitudes are in terms of numerators
divided by propagators, as in \eqref{eq:colornumdecomp}. These
amplitudes obey symmetry properties. They are invariant under cyclic
permutations of indices. They have symmetry under reversal:
$A(1,2,\ldots,n) = (-1)^n A(n,\ldots,2,1)$. These amplitudes obey the
photon-decoupling identity:
\begin{equation}
\sum_{\sigma \in cyclic} {A(1,\sigma(2,3,\ldots,n))}=0,
\end{equation}
and more generally the Kleiss-Kuijf (KK) relations \cite{KK}.
\begin{equation}
A(1,\left\{\alpha \right\},n,\left\{\beta \right\}) = (-1)^{n_\beta} \sum_{\left\{\sigma\right\}_i \in OP(\left\{\alpha \right\},\left\{\beta^T \right\})} {A(1,\left\{\sigma\right\}_i,n)}.
\end{equation}
Here, $OP(\left\{\alpha \right\},\left\{\beta^T \right\})$ is the set
of permutations which preserve the order of each set. The set
$\left\{\beta^T \right\}$ is $\left\{\beta \right\}$ but with reversed
ordering. Using these identities, we can choose a set of $(n-2)!$
color-ordered amplitudes as a KK-independent basis.

In eq.~(\ref{eq:colornumdecomp}) the tree amplitudes are written in
terms of numerators.  The duality between color and kinematics
immediately implies that we can write the amplitude in terms of the
same number of numerators as independent color factors. In principle,
the $(n-2)!$ amplitudes and $(n-2)!$ numerators in this basis could
uniquely specify each other. However, it turns out that the numerators
are not unique -- they can be shifted by kinematic functions in such a
way that the basis amplitudes remain unchanged. See
refs.~\cite{BCJ,BernDHK10,Tye10} for details. This freedom is a residual
generalized gauge freedom, or ``residual freedom''. In other words,
the $(n-2)!$ equations for the basis amplitudes in terms of basis
numerators are non-invertible. Only $(n-3)!$ of the $(n-2)!$
numerators are independent, resulting in $(n-2)!-(n-3)!$ amplitude
relations, known as BCJ amplitude relations.

\subsection{The three-algebra formulation of ABJM theory}

\begin{figure}[tb]
\begin{center}
\includegraphics[width=\textwidth]{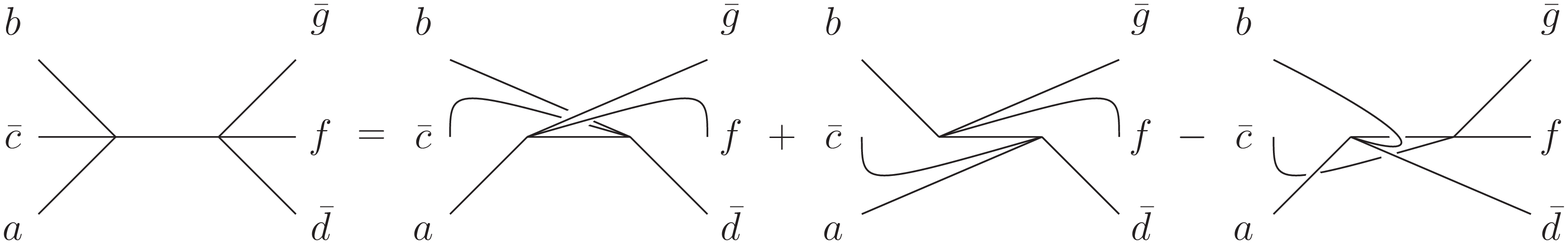}
\end{center}
\caption{\small 
Diagrammatic representation of the generalized Jacobi relation.
}
\label{JacRelationFigure}
\end{figure}

We will consider ABJM theory, a theory of M2 branes
\cite{ABJM}. This theory admits a natural Lie
three-algebra formulation analogous to the Lie two-algebra defining
the gauge symmetry of Yang-Mills theory. ABJM theory is a Chern-Simons-matter
gauge theory with gauge-group $SU(N) \times SU(N)$. Matter
fields are bi-fundamental, and map between the two gauge groups: they are written in terms of group elements
$(T^a)^{b \bar{c}} : V_1 \rightarrow V_2$ where $V_1$ and $V_2$ are
the vector spaces of group elements from the first and second $SU(N)$ groups. Indices are barred to distinguish between the two vector
spaces. The index $a$ is an adjoint index, while indices $b, \bar{c}$ are fundamental and anti-fundamental. We do not use different notation for adjoint and (anti-)fundamental indices as we often show only the adjoint indices. The reader should keep in mind that multiplication of group elements involves contracting fundamental and anti-fundamental indices. The group elements $T^a$ are related by a triple product:
\begin{align}
[T^a,T^b;\bar{T}^{\bar{c}}] &= {f^{a b \bar{c}}}_d T^{d}, &\quad [T^a,T^b;\bar{T}^{\bar{c}}] &= T^a \bar{T}^{\bar{c}} T^b - T^b \bar{T}^{\bar{c}} T^a.
\end{align}
Indices are raised and lowered with the metric $\text{Tr}(T^a \bar{T}^{\bar{b}}) = h^{a \bar{b}}$. The unbarred and barred indices are antisymmetric amongst each other separately, but not together: they are adjoint indices in two different gauge groups. For example, $f^{ab\bar{c} \bar{d}} = - f^{ba\bar{c} \bar{d}} =f^{ba\bar{d} \bar{c}}$. One can write $f^{a c \bar{b} \bar{d}}$ or $f^{a \bar{b} c \bar{d}}$ as they are the same. The structure constants obey a generalized four-term Jacobi relation illustrated in fig.~\ref{JacRelationFigure}:
\begin{equation}
{f^{a b \bar{c}}}_e {f^{e f \bar{g}}}_d = {f^{a f \bar{g}}}_e {f^{e b \bar{c}}}_d+{f^{b f \bar{g}}}_e {f^{a e \bar{c}}}_d-{f_{\bar{e}}}^{f \bar{g} \bar{c}} {f^{a b \bar{e}}}_d.
\label{eq:Jac}
\end{equation}
ABJM amplitudes are only nonzero for even numbers of external
particles due to the theory's bi-fundamental nature. 

The color-ordering and numerator decomposition of ABJM amplitudes proceed along the same lines as in Yang-Mills
theory. The ABJM numerator decomposition takes the same form \eqref{eq:colornumdecomp}, but the color factors are now products of the
four-index structure constants and the sum runs over diagrams with only quartic vertices. Unlike in Yang-Mills theory, generic numerators in the three-algebra formulation of ABJM theory are non-local. The structure constant has four indices, but the matter is coupled to the gauge field by a cubic interaction. The propagators in the color-kinematic decomposition of ABJM amplitudes \eqref{eq:colornumdecomp} specify a graph with four-point interactions, and so some of the propagators that specify three-point interactions must be absorbed into the numerator. The color factors in the numerator decomposition can be expanded into the trace over strings of generators, schematically $\text{Tr}(T^a \bar{T}^{\bar{b}}\ldots)$, by using the following identities:
\begin{align}
&f^{a b \bar{c} \bar{d}} = \text{Tr}([T^a \bar{T}^{\bar{c}} T^b - T^b \bar{T}^{\bar{c}} T^a] \bar{T}^{\bar{d}}),
\\
&{({T^a})^i}_{\bar{j}} \protect\phantom{|} {(\bar{T}^{\bar{b}})^{\bar{m}}}_n \protect\phantom{|} h_{a \bar{b}} =
\delta^{\bar{m}}_{\bar{j}} \protect\phantom{|} \delta^i_n.
\end{align}
The numerator decomposition can therefore be converted into a sum over color-ordered amplitudes, just as in Yang-Mills theory. Color-ordered $2 m$-point amplitudes are defined by
\begin{align}
{\cal A}(\bar{1},2,\bar{3},\ldots~,\overline{2 m-1}, 2 m) =  \sum_{\sigma \in S'_{2m}/Z_{2m} } &\text{Tr}(\bar{T}^{\bar{\sigma}_1}T^{\sigma_2} \bar{T}^{\bar{\sigma}_3}\ldots~\bar{T}^{\bar{\sigma}_{2m-1}} T^{\sigma_{2 m}})  \nonumber
\\
\times & A(\bar{\sigma}_1, \sigma_2,\bar{\sigma}_3,\ldots,~\bar{\sigma}_{2m -1} \sigma_{2 m}). 
\end{align} 
The set $S'_{2m}/Z_{2m}$ is all orderings that have alternating barred and unbarred legs and are unrelated by cyclic permutation. The color-ordered amplitudes can be written in terms of numerators. One color factor is the sum of different trace-strings with different signs, and these signs determine the relative signs of the numerators in the color-ordered amplitudes. For example, consider a color factor $c_i \equiv f^{13\bar{2}\bar{4}} = \text{Tr}([T^1 \bar{T}^{\bar{2}} T^3 - T^3 \bar{T}^{\bar{2}} T^1] \bar{T}^{\bar{4}})$. The associated numerator $n_i$ will enter $A(1 \bar{2} 3 \bar{4})$ and $A(3 \bar{2} 1 \bar{4})$, but with opposite signs. The color-ordered ABJM amplitudes have symmetries similar to those of Yang-Mills amplitudes upon inversion and cyclic permutation. For amplitudes with external bosonic states
\begin{align}
A_{2m}(\bar{1},2,\bar{3},\ldots,\overline{2m-1},2m)&=A_{2m}(\bar{3},4,\ldots,\overline{2m-1},2m,\bar{1},2),\\
A_{2m}(\bar{1},2,\bar{3},\ldots,\overline{2m-1},2m)&=(-1)^{2m-1}A(\bar{1},2m,\overline{2m-1},\ldots,\bar{3},2).
\end{align}
The amplitudes obey KK-type identities, though these identities are not entirely understood beyond six points~\cite{HuangJL13}. Such identities are linear relations between amplitudes with integer coefficients, just as in Yang-Mills theory. In the following, we sometimes denote barred and unbarred indices by even and odd particle labels. We do not need to keep
track of ordering between the two types of indices. For a more detailed review of ABJM theory's three-algebra formulation, see
refs. \cite{HuangJL13, Elvang13}.

\section{The color-kinematic duality in ABJM theory}
Evidence for the color-kinematic duality in ABJM theory was recently found in
tree-level scattering amplitudes \cite{Bargheer12}. Testing the
duality in ABJM theory proceeds just as in Yang-Mills theory: requiring the
numerators satisfy the duality between kinematics and color generates
BCJ amplitude identities, which can be verified by using the explicit
amplitudes. Details of how to calculate the explicit amplitudes, as
well as some lower-point examples, are described in
ref. \cite{Gang11}.

We describe the color-kinematic duality at four, six, and
eight points. The color-kinematic duality is trivially satisfied at
four points. At six and eight points, we consider amplitudes with
bosonic external states. The six-point case is the first non-trivial
instance of the duality \cite{HuangJL13}. Next, we explicitly demonstrate the
duality for six points as a warmup to our work at eight
points. Duality-satisfying six-point numerators in terms of amplitudes
are provided, as these do not appear elsewhere in the literature. At
eight points, we show how to construct the amplitudes in terms of
numerators. Numerical analysis shows that the numerators satisfy the
color-kinematic duality but do not have any residual freedom -- they
are uniquely specified by amplitudes. We give explicit expressions for
the generalized Jacobi identities the numerators satisfy and the
eight-point amplitudes in terms of numerators in the attached files
online~\cite{Online}. All our analysis is for three dimensional on-shell momenta, consistent
with the space-time dimension of the theory.
As ref. \cite{HuangJ12} found, implications of the color-kinematic
duality change when these conditions are relaxed: when momenta are off-shell or 
taken in more than three dimensions, the freedom that produces the BCJ amplitude
relation at six points is no longer present.

\subsection{Four points}
\begin{figure}[htb]
\begin{center}
\includegraphics[scale=.5]{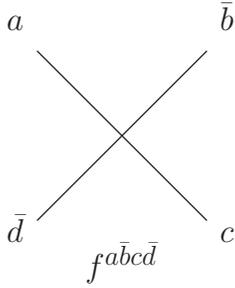}
\end{center}
\caption{\small 
The four-point diagram with its associated color factor. This diagram is the elementary vertex of ABJM theory in the three-algebra formulation. 
\label{FourPointFigure}
}
\end{figure}

In the three-algebra formulation of ABJM theory, the vertex associated with
$f^{a \bar{b} c \bar{d}}$ comes from a four-point diagram, illustrated
in fig.~\ref{FourPointFigure}, with a non-local numerator. At four
points, there is only one independent color-ordered amplitude after
accounting for amplitude symmetries.  When assembling the indices on a
structure constant, one must choose a convention: whether to begin
with an unbarred or barred index, and whether to move clockwise or
counter-clockwise. We chose to begin from an unbarred index and move
clockwise in the diagram. We also use the convention that all momenta are incoming.
With these conventions the
four-point superamplitude is \cite{Gang11}
\begin{equation}
A(1,\bar{2},3,\bar{4}) = \frac{\delta^{(3)}(P) \delta^{(6)}(Q) }{\left\langle 14\right\rangle \left\langle 34\right\rangle} f^{a_1 \bar{a}_2 a_3 \bar{a}_4}.
\end{equation}
The delta functions conserve momentum $P^{\alpha \beta} =
\sum_i{p_i^{\alpha \beta}}$ and supermomentum $Q^{\alpha I} =
\sum_i{q_i^{\alpha I}}$. As we have $\mathcal{N} = 6$ real supercharges, these can be grouped into 3 complex Grassmann-valued spinors.
We have $q_i^{\alpha I} = \lambda_i^\alpha \eta^I$ for the $i$-th particle. The label $I$ is the index for the $SO(6)$ R-symmetry, and $\alpha$ labels the supercharge number, running from one to three \cite{Gang11,Elvang13}. The amplitude is written using spinor-helicity formalism (see e.g. \cite{ManganoReview}).

\subsection{Six points}

\begin{figure}[htb]
\begin{center}
\includegraphics[scale=.5]{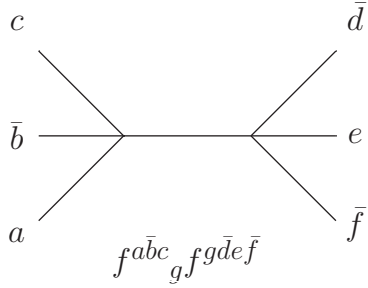}
\end{center}
\caption{\small The six-point diagram with its associated color
  factor.  This diagram does not have left-right reflection symmetry.
\label{SixPointFigure}
}
\end{figure}

At six points, the color-kinematic duality holds
\cite{Bargheer12,HuangJ12}. We present explicit formulas showing how
it holds.  The six-point color factors are products of two structure
constants, and obey the generalized Jacobi identity
\eqref{eq:Jac}. The full amplitude has nine independent channels:
\begin{equation}
\mathcal{A}(1,2,3,4,5,6)=\frac{c_1 n_1}{s_{123}}+\frac{c_2 n_2}{s_{156}}+\frac{c_3 n_3}{s_{126}}+\frac{c_4 n_4}{s_{134}}+\frac{c_5 n_5}{s_{125}}+\frac{c_6 n_6}{s_{124}}+\frac{c_7 n_7}{s_{136}}+\frac{c_8 n_8}{s_{145}}+\frac{c_9 n_9}{s_{146}}.
\end{equation}
The kinematic invariants are $s_{ijk} = (p_i+p_j+p_k)^2$. The pole
structure specifies the diagram, and so fixes the color factor up to
an overall sign:
\begin{alignat}{3}
c_1&={f^{123}}_a f^{a456}, &\quad\quad c_2 &={f^{561}}_a f^{a234}, &\quad\quad c_3&={f^{345}}_a f^{a612}, \nonumber
\\
c_4&={f^{134}}_a f^{a562}, &\quad\quad c_5&={f^{521}}_a f^{a436}, &\quad\quad c_6&={f^{365}}_a f^{a124}, \nonumber
\\
c_7&={f^{163}}_a f^{a254}, &\quad\quad c_8&={f^{541}}_a f^{a632}, &\quad\quad c_9&={f^{325}}_a f^{a416}.
\end{alignat}
There are five amplitudes independent under the KK-type relations. We choose the following amplitudes as our basis amplitudes: 
\begin{alignat*}{2}
A(1,2,3,4,5,6)&=\frac{n_1}{s_{123}}+\frac{n_2}{s_{156}}+\frac{n_3}{s_{126}}, &\quad\quad
A(1,4,3,6,5,2)&=\frac{n_4}{s_{134}}+\frac{n_5}{s_{125}}+\frac{n_6}{s_{124}}, 
\\
A(1,6,3,2,5,4)&=\frac{n_7}{s_{136}}+\frac{n_8}{s_{145}}+\frac{n_9}{s_{146}}, &\quad\quad
A(1,4,3,2,5,6)&=-\frac{n_4}{s_{134}}-\frac{n_2}{s_{156}}-\frac{n_9}{s_{146}}, 
\end{alignat*}
\begin{equation}
A(1,6,3,4,5,2)=-\frac{n_7}{s_{136}}-\frac{n_5}{s_{125}}-\frac{n_3}{s_{126}}.
\end{equation}
The relative sign of each numerator is conveniently
 determined by switching to the trace expansion of its color factor.
 Next, we require that the numerators satisfy the duality between color and kinematics. Each numerator must obey the same identities as its sibling color factor:
\begin{alignat*}{2}
n_5&=-n_2+n_3+n_4, &\quad\quad n_6&=n_1-n_2+n_4,
\\
n_8&=n_1-n_3+n_7, &\quad\quad n_9&=n_2-n_3+n_7.
\end{alignat*}
We can choose $\left\{n_1, n_2, n_3, n_4, n_7\right\}$ as a basis independent under the color-kinematic duality. There are now five independent numerators, the same number as the KK-independent amplitudes. Our set of KK-independent amplitudes are now
\begin{align}
A(1,2,3,4,5,6)&\equiv A_1 = \frac{n_1}{s_{123}}+\frac{n_2}{s_{156}}+\frac{n_3}{s_{126}} \nonumber
\\
A(1,4,3,6,5,2)&\equiv A_2 = \frac{n_4}{s_{134}}+\frac{-n_2+n_3+n_4}{s_{125}}+\frac{n_1-n_2+n_4}{s_{124}} \nonumber
\\
A(1,6,3,2,5,4)&\equiv A_3 = \frac{n_7}{s_{136}}+\frac{n_1-n_3+n_7}{s_{145}}+\frac{n_2-n_3+n_7}{s_{146}} \nonumber
\\
A(1,4,3,2,5,6)&\equiv A_4 = -\frac{n_4}{s_{134}}-\frac{n_2}{s_{156}}-\frac{n_2-n_3+n_7}{s_{146}} \nonumber
\\
A(1,6,3,4,5,2)&\equiv A_5 = -\frac{n_7}{s_{136}}-\frac{-n_2+n_3+n_4}{s_{125}}-\frac{n_3}{s_{126}}.
\label{eq:6ptpostjac}
\end{align}

We can solve for the numerators one by one and find that one numerator drops out of our equations. To be concrete, we begin by solving for $n_1$: choosing one of the amplitude-numerator equations above and solving for $n_1$ in terms of an amplitude and the remaining numerators, we substitute this expression for $n_1$ into the remaining amplitude equations. The equations now relate the five amplitudes to four numerators. We solve for $n_2$ and $n_3$ in the same way, leaving two equations that relate the five amplitudes to $n_4$ and $n_7$. When solving for $n_4$ in one equation and substituting the result into the remaining equation, we find that $n_7$ drops out. What remains is an equation relating the five amplitudes -- this is the BCJ amplitude relation. One of the five numerators is arbitrary. In other words, the coefficient matrix of the numerators in the amplitude equations has rank four, while there are five numerators. This is similar to the situation in Yang-Mills theory~\cite{BCJ}. 

We may therefore choose one numerator to have an arbitrary value. As the remaining numerators depend on this numerator, 
the remaining numerators depend on our choice. A convenient choice is to simply set $n_7$ to zero
and solve for the remaining numerators in terms of amplitudes. Imposing 
three-dimensional momentum conservation produces a lengthy expression, so we display the
solution using the original kinematic invariants:
\begin{alignat}{2}
n_1 &= s_{123} \left(A_1 + \frac{A}{s_{234}}+ \frac{E}{D} \left(\frac{1}{s_{345}}-\frac{B}{C} \frac{1}{s_{234}}\right)\right), &\quad
n_2 &= -A+\frac{B E}{C D} ,\nonumber
\\
n_4 &= s_{134}\left(-A_4 +\left(A-\frac{B E}{C D}\right)\left(\frac{1}{s_{234}}+\frac{1}{s_{235}}\right)-\frac{E }{D} \frac{1}{ s_{235}}\right), &\quad
n_3 &= -\frac{E}{D}.
\end{alignat}
The quantities $A,B,C,D,E$ have been defined for convenience and are
\begin{align}
A=&\frac{-\frac{A_3}{s_{123}}+\frac{A_1}{s_{236}}}{\frac{1}{s_{123} s_{235}}-\frac{1}{s_{234} s_{236}}} \nonumber
\\
B=&-\frac{\frac{1}{s_{235}}+\frac{1}{s_{236}}}{s_{123}}-\frac{1}{s_{236} s_{345}}\nonumber
\\
C=&\frac{1}{s_{123} s_{235}}-\frac{1}{s_{234} s_{236}}\nonumber
\\
D=&-B \left(-\frac{-\left(\frac{1}{s_{234}}+\frac{1}{s_{235}}\right) \left(-\frac{1}{s_{134}}-\frac{1}{s_{346}}-\frac{1}{s_{356}}\right)+\frac{\frac{1}{s_{346}}+\frac{1}{s_{356}}}{s_{134}}}{s_{123}}-\frac{1}{s_{134} s_{234} s_{356}}\right) \nonumber
\\
&+C \left(-\frac{-\frac{1}{s_{134} s_{346}}+\frac{-\frac{1}{s_{134}}-\frac{1}{s_{346}}-\frac{1}{s_{356}}}{s_{235}}}{s_{123}}-\frac{1}{s_{134} s_{345} s_{356}}\right) \nonumber
\\
E=& \protect\phantom{|} C\left(-\frac{\frac{A_2}{s_{134}}-A_4 \left(-\frac{1}{s_{134}}-\frac{1}{s_{346}}-\frac{1}{s_{356}}\right)}{s_{123}}+\frac{A_1}{s_{134} s_{356}}\right. \nonumber
\\
&+\left.A\left(\frac{-\left(\frac{1}{s_{234}}+\frac{1}{s_{235}}\right) \left(-\frac{1}{s_{134}}-\frac{1}{s_{346}}-\frac{1}{s_{356}}\right)+\frac{\frac{1}{s_{346}}+\frac{1}{s_{356}}}{s_{134}}}{s_{123}}+\frac{1}{s_{134} s_{234} s_{356}}\right)\right).
\end{align}

We now substitute the solutions for ${n_1,n_2,n_3,n_4}$ into the
expression for $A_5$ in \eqref{eq:6ptpostjac}. Since $n_7$ drops out,
it leaves behind a single nontrivial BCJ amplitude relation between
the five partial amplitudes. We have confirmed that this relation
holds numerically in the actual amplitudes by plugging in explicit values for the amplitudes, which were obtained in ref. \cite{Gang11}. The
six-point case is discussed further in ref. \cite{HuangJL13}, which
gives the BCJ amplitude relation and checks it for the
superamplitude. While in Yang-Mills theory the freedom to adjust numerators 
may be used to keep the color-kinematic numerators local, in the three-algebra formulation 
of ABJM theory numerators are inherently non-local.
The freedom to adjust $n_7$ cannot be used to make the remaining numerators local functions.

\subsection{Eight points}

\begin{figure}[htb]
\begin{center}
\includegraphics[width=\textwidth]{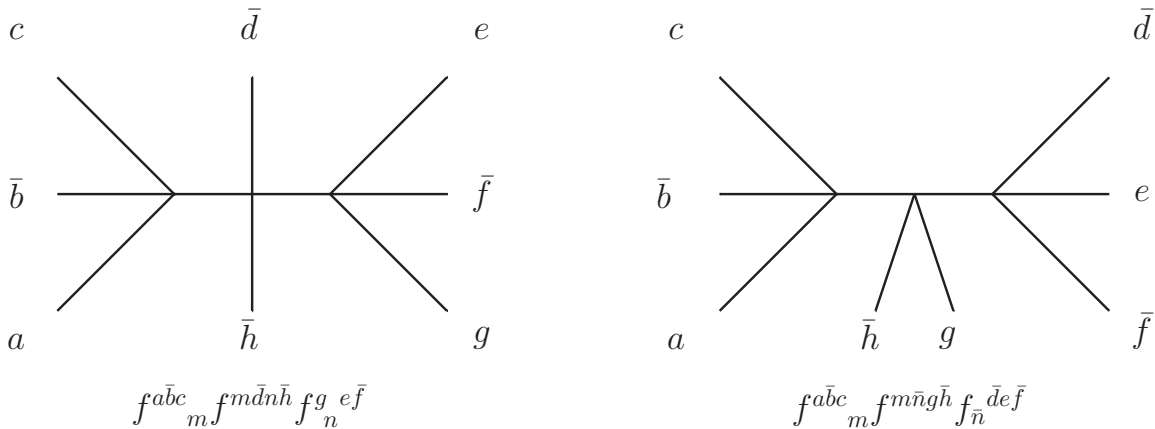}
\end{center}
\caption{\small The two diagrams that contribute numerators to the
  ABJM color-stripped amplitude.  The left diagram has a color factor
  of the form ${f^{a \bar{b} c}}_m f^{m \bar{d} n \bar{h}} f^{g
    \protect\phantom{a} e \bar{f}}_{\protect\phantom{a} n}$. The
  external indices in the middle are both barred or, for a different
  leg ordering than shown above, both unbarred. The right diagram has
  a color factor of the form ${f^{a \bar{b} c}}_m f^{m \bar{n} g
    \bar{h} }f_{\bar{n}} \protect\phantom{}^ {\bar{d} e \bar{f}}$. One
  middle index is barred while the other is not. 
 \label{EightPointFigure}
}
\end{figure}

The eight-point case proceeds in the same fashion as the six-point
case. The two distinct diagrams that contribute are shown in
fig.~\ref{EightPointFigure}. For clarity, we provide one color-stripped
amplitude from which all others are obtained by relabelling. In this
amplitude, we label each numerator by the color factor it is
associated with, suppressing the summed indices on the structure
constants. For example, we denote the numerator associated with the
color factor ${f^{1 \bar{2} 3}}_a f^{a \bar{4}
  \bar{8} b} f^{5 7 \phantom{a} \bar{6}}_{\phantom {aa} b}$ as
$n_{123,48,576}$. In assigning the color-factors, we use the
convention that the summed indices of the color factors are as close
together as possible, and the right-most structure constant has an
index which is higher than the left-most structure constant's
indices. We suppress the bars and take odd and even labels to correspond to
the barred and unbarred indices separately.  The color-stripped amplitude then is
\begin{align}
A(1,2,3,4,5,6,7,8)=
&-\frac{n_{132,48,576}}{s_{123}s_{567}}-\frac{n_{354,62,718}}{s_{345}s_{781}}
+\frac{n_{324,51,768}}{s_{234}s_{678}}+\frac{n_{546,73,182}}{s_{456}s_{812}} \nonumber
\\
&-\frac{n_{132,78,546}}{s_{123}s_{456}}-\frac{n_{354,12,768}}{s_{345}s_{678}} 
-\frac{n_{576,34,182}}{s_{567}s_{812}}-\frac{n_{718,56,324}}{s_{781}s_{234}} \nonumber
\\
&+\frac{n_{576,18,324}}{s_{567}s_{234}}+\frac{n_{718,32,546}}{s_{781}s_{456}}
+\frac{n_{132,54,768}}{s_{123}s_{678}}+\frac{n_{354,76,182}}{s_{345}s_{812}}.
\end{align}
All other color-stripped eight-point amplitudes can be obtained from this expression by
relabelling. The signs of each numerator depends on the trace string's
contribution to each color factor, similar to the Yang-Mills
case~\cite{BCJ}. The
eight-point amplitude has 216 different color factors, and therefore
216 corresponding numerators. We have checked that our numerator representation of
the color-stripped amplitudes obey the correct cyclic permutation
properties and KK-type identities as listed in
ref. \cite{HuangJL13}. We repeat these symmetries below, using
the bar-unbar notation for clarity. 
The symmetries are
\begin{alignat}{2}
\text{Cyclic shift by $2m$:}& &\quad A(1 \bar{2} 3 \bar{4} 5 \bar{6} 7 \bar{8}) &= A( 7 \bar{8}1 \bar{2} 3 \bar{4} 5 \bar6),
\\
\text{Reversal:}& &\quad A(1 \bar{2} 3 \bar{4} 5 \bar{6} 7 \bar{8}) &= -A(1 \bar{8} 7 \bar{6} 5 \bar{4} 3 \bar{2}),
\end{alignat}
where $m$ is an arbitrary integer.
An example of an eight-point KK-type identity satisfied by our amplitudes is
\begin{align}
&-A(1 2 3 4 5 8 7 6)-A(1 4 3 2 5 8 7 6)+ A(1 6 3 8 5 2 7 4)+A(1 6 3 8 5 4 7 2) \nonumber
\\
&+A(1 6 7 8 3 2 5 4)+A(1 2 7 6 3 8 5 4)+A(1 4 7 6 3 8 5 2)+A(1 6 7 8 3 4 5 2) \nonumber
\\
&+A(1 6 7 2 3 8 5 4)+A(1 6 7 4 3 8 5 2)+A(1 6 3 2 7 8 5 4)+A(1 6 3 4 7 8 5 2) \nonumber
\\
&+A(1 6 3 8 7 2 5 4)+A(1 6 3 8 7 4 5 2)+A(1 4 3 6 7 8 5 2)+A(1 2 3 6 7 8 5 4)=0.
\end{align}
There are 57 amplitudes independent under the KK-type identities. We call these the KK amplitudes. The basis was chosen by eliminating all linear dependence between amplitudes written in terms of numerators. As in Yang-Mills theory, we know the resulting linear amplitude relations are not BCJ amplitude relations, because no generalized Jacobi relations have been used. Eliminating the linear dependence between amplitudes in this step corresponds to exhausting the KK-type amplitude relations. No closed form for the KK-type identities have been found, but the number of basis amplitudes agrees with the result in ref. \cite{HuangJL13}. We choose a set of KK amplitudes and present a sample below. The full set of equations is contained in an attached file online~\cite{Online}.
\begin{align*}
A(12345876)=&-\frac{n_1}{s_{123} s_{876}}-\frac{n_2}{s_{123} s_{587}}-\frac{n_7}{s_{123} s_{458}}+\frac{n_{28}}{s_{587} s_{612}}-\frac{n_{30}}{s_{458} s_{612}}-\frac{n_{36}}{s_{345} s_{612}}
\\&-\frac{n_{136}}{s_{458} s_{761}}-\frac{n_{139}}{s_{345} s_{761}}+\frac{n_{144}}{s_{234} s_{761}}+\frac{n_{163}}{s_{234} s_{876}}+\frac{n_{164}}{s_{234} s_{587}}+\frac{n_{208}}{s_{345} s_{876}},
\\
A(12365478)=&\frac{n_5}{s_{123} s_{478}}-\frac{n_8}{s_{123} s_{547}}+\frac{n_9}{s_{123} s_{654}}+\frac{n_{48}}{s_{547} s_{812}}-\frac{n_{49}}{s_{654} s_{812}}-\frac{n_{51}}{s_{365} s_{812}}
\\&+\frac{n_{154}}{s_{654} s_{781}}-\frac{n_{155}}{s_{365} s_{781}}+\frac{n_{160}}{s_{236} s_{781}}-\frac{n_{171}}{s_{236} s_{478}}+\frac{n_{173}}{s_{236} s_{547}}-\frac{n_{212}}{s_{365} s_{478}},
\\
A(12365874)=&\frac{n_2}{s_{123} s_{587}}-\frac{n_3}{s_{123} s_{658}}-\frac{n_5}{s_{123} s_{874}}+\frac{n_{11}}{s_{412} s_{587}}-\frac{n_{12}}{s_{412} s_{658}}-\frac{n_{18}}{s_{365} s_{412}}
\\&-\frac{n_{100}}{s_{658} s_{741}}-\frac{n_{103}}{s_{365} s_{741}}+\frac{n_{107}}{s_{236} s_{741}}+\frac{n_{170}}{s_{236} s_{587}}+\frac{n_{171}}{s_{236} s_{874}}+\frac{n_{212}}{s_{365} s_{874}}.
\end{align*}
The four-term generalized Jacobi relations the numerators must satisfy
according to the color-kinematic duality are also listed in an
attached file online. These relations are solved, and all numerators
are specified by the 57 basis numerators. After imposing the Jacobi
relations, the 57 amplitude equations are expressed in terms of the 57 basis
numerators. Here, the number of KK amplitudes and numerators independent under the generalized Jacobi relations are the same, just as in Yang-Mills theory.
However, the surprise here is that these equations turn out to be invertible. 
Solving the system analytically is difficult, but it is straightforward to check invertibility numerically by
using explicit values for the three-dimensional momentum that obey momentum conservation. In other words,
the numerators have no residual freedom: each numerator is uniquely
specified as a function of gauge-invariant amplitudes. Without
residual freedom, the eight-point amplitudes have no BCJ amplitude
relations. The lack of BCJ amplitude relations is a surprising result.

At eight points, the color-kinematic duality is automatically satisfied. Through the numerator decomposition the amplitudes obey the KK-type identities manifestly, and unlike in Yang-Mills theory, there are no BCJ amplitudes relations to check.

\section{Discussion}

We have examined bosonic amplitudes through eight points in ABJM theory and found that
the duality between color and kinematics holds, but at eight 
points attendant BCJ amplitude relations are not present. The color-kinematic duality
specifies the eight-point numerators as unique functions of
gauge-invariant partial amplitudes. We also have given explicit expressions
necessary for analysis of ABJM amplitudes that do not appear in the
literature. At six points, we have solved for a basis of
color-kinematic numerators in terms of amplitudes. At eight points, we
presented a set of amplitudes independent under the Kleiss-Kuijf relations in terms of numerators, as well as the
relations the numerators satisfy according to the color-kinematic
duality. The full eight-point expressions can be found in 
an ancillary online file~\cite{Online}.

The eight-point amplitude-numerator decomposition we provided can
be used to study possible  relations between ABJM amplitudes, 
or equivalently between ABJM numerators \cite{HuangJL13}.
A recent twistor string construction for ABJM theory may also provide insight \cite{Engelund14}.
BLG theory is a special case of ABJM theory, and possesses the
color-kinematic duality, residual freedom, and the double-copy property
\cite{HuangJL13}. The mapping between BLG and ABJM amplitudes is straightforward at four and six
points, but is not fully understood at eight points \cite{HuangJL13} and needs further investigation.
While in ABJM theory the double-copy procedure leads to 
supergravity amplitudes at lower points,
at eight points the double-copy property fails
\cite{HuangJL13}.  One might suspect that there is a connection between the simultaneous
disappearances of the double-copy property and residual freedom. Further study of the eight point ABJM amplitudes may provide insight into the double-copy property through side-by-side comparison with BLG amplitudes. 

Using the data we have provided, further analysis of the eight-point
ABJM amplitudes may provide important clues about the
color-kinematic duality, residual freedom, and the double-copy
property. The presence of the color-kinematic duality in ABJM theory without
associated BCJ amplitude relations emphasizes the basic role 
numerators play in the duality compared to the amplitude relations.
The eight-point ABJM amplitudes are the lowest-point
amplitudes that allow the color-kinematic duality without possessing residual freedom
to rearrange the numerators or 
yielding gravity amplitudes via the double-copy property. 
These amplitudes therefore provide an interesting avenue for further understanding the 
role of these properties in the duality between color and kinematics.

\section{Acknowledgements}

We thank Z. Bern and Y.-t. Huang for guidance and discussions; and
Y.-t. Huang for expertise and assistance concerning three-dimensional Chern-Simons-matter theories. We also wish to thank
S. Davies, M. Gutperle, H. Johansson, P. Kraus, S. Litsey, J. Nohle, and
J. Stankowicz for helpful discussions and comments.

\end{document}